\PassOptionsToPackage{bookmarks={false}}{hyperref}
\documentclass[conference]{IEEEtran}
\IEEEoverridecommandlockouts
\usepackage{cite}
\usepackage{amsmath,amssymb,amsfonts}
\usepackage{algorithmic}
\usepackage[ruled,vlined]{algorithm2e}
\usepackage{amsfonts}
\usepackage{graphicx}
\usepackage{textcomp}
\usepackage{xcolor}
\newcommand*{\affaddr}[1]{#1}
\newcommand*{\affmark}[1][*]{\textsuperscript{#1}}
\def\BibTeX{{\rm B\kern-.05em{\sc i\kern-.025em b}\kern-.08em
    T\kern-.1667em\lower.7ex\hbox{E}\kern-.125emX}}
\usepackage[T1]{fontenc}
\usepackage[utf8]{inputenc}
\usepackage{authblk}
\usepackage{tabularx}
\usepackage{threeparttable}
\usepackage{verbatim} 
\usepackage{hyperref}

\usepackage{tikz}
\usepackage{textcomp}
\usepackage{lipsum}

\newcommand\copyrighttext{%
  \footnotesize \textcopyright 2020 IEEE. Personal use of this material is permitted.
  Permission from IEEE must be obtained for all other uses, in any current or future 
  media, including reprinting/republishing this material for advertising or promotional 
  purposes, creating new collective works, for resale or redistribution to servers or 
  lists, or reuse of any copyrighted component of this work in other works. }
\newcommand\copyrightnotice{%
\begin{tikzpicture}[remember picture,overlay]
\node[anchor=south,yshift=10pt] at (current page.south) {\fbox{\parbox{\dimexpr\textwidth-\fboxsep-\fboxrule\relax}{\copyrighttext}}};
\end{tikzpicture}%
}

\begin{document}

\title{TSception:A Deep Learning Framework for Emotion Detection Using EEG}

\author{Yi Ding\affmark[1],
   \hspace{2mm}
   Neethu Robinson\affmark[1],
   \hspace{2mm}
   Qiuhao Zeng\affmark[1],
   \hspace{2mm}
   Duo Chen\affmark[1],
   \hspace{2mm}
   Aung Aung Phyo Wai\affmark[1],
   \hspace{2mm}
   \\
   Tih-Shih Lee\affmark[2,3],
   \hspace{2mm}
   Cuntai Guan\affmark[1]\affmark[*]
   \hspace{2mm}
   \\
   \affaddr{\affmark[1]School of Computer Science and Engineering, Nanyang Technological University, Singapore} \\
   \affaddr{\affmark[2]Neuroscience and Behavioral Disorders Program,\\ Duke University - National University of Singapore Medical School, Singapore, Singapore} \\
   \affaddr{\affmark[3]Singapore General Hospital, Singapore, Singapore} \\
   \affmark[1]\{ding.yi, nrobinson, qiuhao.zeng, chenduo, apwaung, ctguan\}@ntu.edu.sg,\\ \affmark[2, 3]tihshih.lee@duke-nus.edu.sg
   \thanks{* Cuntai Guan is the Corresponding Author.}
}

\maketitle

\begin{abstract}
In this paper, we propose a deep learning framework, TSception, for emotion detection from electroencephalogram (EEG). TSception consists of temporal and spatial convolutional layers, which learn discriminative representations in the time and channel domains simultaneously. The temporal learner consists of multi-scale 1D convolutional kernels whose lengths are related to the sampling rate of the EEG signal, which learns multiple temporal and frequency representations. The spatial learner takes advantage of the asymmetry property of emotion responses at the frontal brain area to learn the discriminative representations from the left and right hemispheres of the brain. In our study, a system is designed to study the emotional arousal in an immersive virtual reality (VR) environment. EEG data were collected from 18 healthy subjects using this system to evaluate the performance of the proposed deep learning network for the classification of low and high emotional arousal states. The proposed method is compared with SVM, EEGNet, and LSTM. TSception achieves a high classification accuracy of 86.03\%, which outperforms the prior methods significantly ($p$<0.05).
\end{abstract}

\copyrightnotice
\section{Introduction}
Emotions are fundamental in the daily life of human beings. Emotions can be mapped into the Valence, Arousal, and Dominance (VAD) dimensions\cite{7946165}. Among three dimensions, Emotional Arousal (EA) detection plays an important role in the diagnosis and therapy of psychological disabilities, such as anxiety disorder\cite{MORENA201659}\cite{info:doi/10.2196/13869}, autistic spectrum disorders (ASD)\cite{doi:10.1002/hbm.23041}. Research\cite{lane_ryan_nadel_greenberg_2015} on emotion-focused therapy (EFT) demonstrated that the emotional arousal is critical to psychotherapeutic success. However the detection of the emotional arousal is still a challenging task for the human-machine interaction system.

Brain computer interface (BCI) system enables the computer to perceive the arousal mental state of the human, using machine learning and signal processing technology\cite{10.1371/journal.pone.0213516}. Electroencephalography (EEG) is collected by several electrodes located on the surface of the human head, reflecting the potential neural activity directly. Compared to other emotional signals, such as facial expression and natural language, EEG contains more comprehensive information regarding human mental state and objective evaluation. Emotional arousal detection by EEG contains three main parts: pre-processing, feature extraction and classifier training. The artifact and noise (e.g. eyes blinks, 60 Hz noise) will be removed during the pre-processing stage. Power Spectral Density (PSD) of different frequency bands, Differential Entropy (DE), Event-Related De/Synchronizations (ERD/ERS), Event-Related Potentials
(ERP), etc. are commonly extracted as features. A set of features is then selected to train a classifier. 

A lot of research work has been conducted to solve the EEG emotional state classification problem\cite{7946165}\cite{Craik_2019}. Atkinson \textit{et al}. \cite{ATKINSON201635} proposed an efficient feature selection method to improve the SVM classifier performance on emotional arousal detection, with the accuracy being 73.14\%. Zheng \textit{et al}\cite{7938737} investigated stable patterns of electroencephalogram (EEG) over time for emotion recognition, using Discriminative Graph Regularized Extreme Learning Machine with DE features. Li \textit{et al.}\cite{8634938} constructed emotion-related brain networks with phase-locking value and adopted a multiple feature fusion approach for emotion recognition. Recently, deep learning methods have shown promising classification performance in BCI, such as motor imagery classification\cite{doi:10.1002/hbm.23730} \cite{8897723}\cite{Tabar_2016}\cite{Lawhern_2018}\cite{8310961}, emotion recognition \cite{7883875}\cite{Li2018}\cite{10.3389/fnins.2018.00162}\cite{8567966}, and mental-task classification\cite{Fahimi_2019}\cite{JIAO2018582}\cite{8607897}. Yang \textit{et. al.}\cite{7883875} designed a hierarchical network structure with sub-network nodes to classify three emotional states. Li \textit{et al.}\cite{Li2018} proposed a Hierarchical Convolutional Neural Networks (HCNN) to extract the spatial information of the EEG electrodes by mapping the EEG signal into a 2D location map. Li \textit{et. al.}\cite{10.3389/fnins.2018.00162} applied 18 kinds of linear and non-linear features to study the cross-subject emotion recognition problems, achieving 59.06\% and 83.33\% on two public datasets. Although many machine learning methods have been proposed for emotional arousal detection, most of them highly rely on the hand-extracted features. Vernon J Lawhern \textit{et al}\cite{Lawhern_2018} proposed EEGNet, an end-to-end deep learning framework that can extract the hidden temporal and spatial patterns from the raw EEG data. 

Inspired by the Inception block of GoogleNet\cite{Szegedy_2015_CVPR}, we proposed TSception, a deep learning framework for EEG signal classification. It uses temporal and spatial learners to learn more discriminative representations for EEG signals in time and space domains simultaneously. There are two types of convolutional learners in TSception: temporal learner and spatial learner. The temporal learner has multi-scale convolutional kernels, learning more discriminative multiple temporal and frequency representations. Psychophysiological evidence\cite{doi:10.1111/psyp.13028} indicates that the left and right halves of the human frontal brain areas differentially associate with particular emotions and affective traits. The spatial learner takes advantage of the frontal area of brain emotional asymmetry, using hemisphere kernels to learn the proper representation of the information from the right and left brain. 

Studies\cite{FELNHOFER201548}\cite{info:doi/10.2196/13869}\cite{8914326} have shown the VR can induce targeted emotion effectively. In order to study the emotional arousal in the immersive VR environment and evaluate the proposed algorithm, we designed a VR-BCI system and collected EEG data from 18 healthy subjects in the VR environment. 

The major contribution of this work can be summarised as: 
\begin{itemize}
\item Designed a new deep learning framework, which uses temporal/spatial learners to learn time/space discriminative EEG representations of low and high emotional arousal states. The convolutional kernels in temporal learner have multi-scale lengths related to the sampling rate, learning multiple temporal and frequency representations parallelly. The spatial learner considers the frontal area of brain emotional asymmetry to learn global-local spatial representations of EEG signals.
\item Designed and implemented the experiment and system to collect emotional arousal EEG data in the VR environment.
\item Finally,  the proposed method is compared with SVM using relative power and DE as features, EEGNet, LSTM, together with two simplified self-comparison versions, namely, Tception and Sception. 
\end{itemize}

\section{Related Work}
\subsection{GoogleNet and Inception Block}
GoogleNet, also known as Inception-V1, won the 2014-ILSVRC competition \cite{Szegedy_2015_CVPR}. In GoogleNet,
small blocks are used instead of conventional convolutional layers. The inception block was introduced in the GoogleNet architecture. In inception blocks, split, transform and merge operations are achieved to improve the learning of varying represents for the same object in different images. The main idea of the Inception block is to extract spatial patterns using multi-scale convolutional kernels (1x1, 3x3, 5x5) on each layer. 

\subsection{Convolutional Neural Network for EEG Data}
Different from images, the EEG data can be treated as 2D time series, whose dimensions are channels (EEG electrodes) and time respectively. The channels in this paper are the EEG electrodes instead of RGB dimensions in images or the input/output channels for convolutional layers. Because the electrodes are located at different areas on the surface of the human's head, the channel dimension contains spatial information of EEG; the time dimension is full of temporal information instead. In order to train a classifier, the EEG signal will be split into shorter time segments by a sliding window with a certain overlap along the time dimension. Each segment will be one input sample for the classifier.

Recently the convolutional neural networks have shown promising results in BCI\cite{8914184}\cite{8008420}\cite{Lawhern_2018}. J. Li \textit{et al}.\cite{Li2018} constructed the EEG data into a sparse 2D map according to the relative location of the electrodes for each time point. Then $(N,N)$ sized convolutional kennels were applied to do convolution. It can capture the local spatial pattern by sharing the kernels step by step but will lose the global spatial information since $N$ is usually smaller than the length of the input. R. T. Schirrmeister \textit{et. al.}\cite{doi:10.1002/hbm.23730} designed a deep ConvNet, which has 1D temporal convolutional kernels and global spatial convolutional kernels to extract temporal and global spatial information from EEG signal. N. Robinson \textit{et. al.}\cite{8914184} presented a deep learning-driven EEG-BCI system to perform decoding of hand motor imagery using deep convolution neural network architecture. Fahimi \textit{et. al.}\cite{Fahimi_2019} use a deep convolutional neural network to build an inter-subject transfer learning framework for attentive mental state detection. Recently, Vernon J Lawhern \textit{et al.}\cite{Lawhern_2018} proposed EEGNet,  which contains the depth-wise convolution kernel of size $(N,1)$.  It can extract global spatial dependency by letting the $N$ equal to the length of the channel dimension. In EEGNet, there are also 1D temporal kernels with a single size in each layer to learn temporal information. 
  
\section{Methodology}
\subsection{General Structure of Proposed Network -  TSception}
\begin{figure*}[htp]
    \centering
    \includegraphics[width= 18cm]{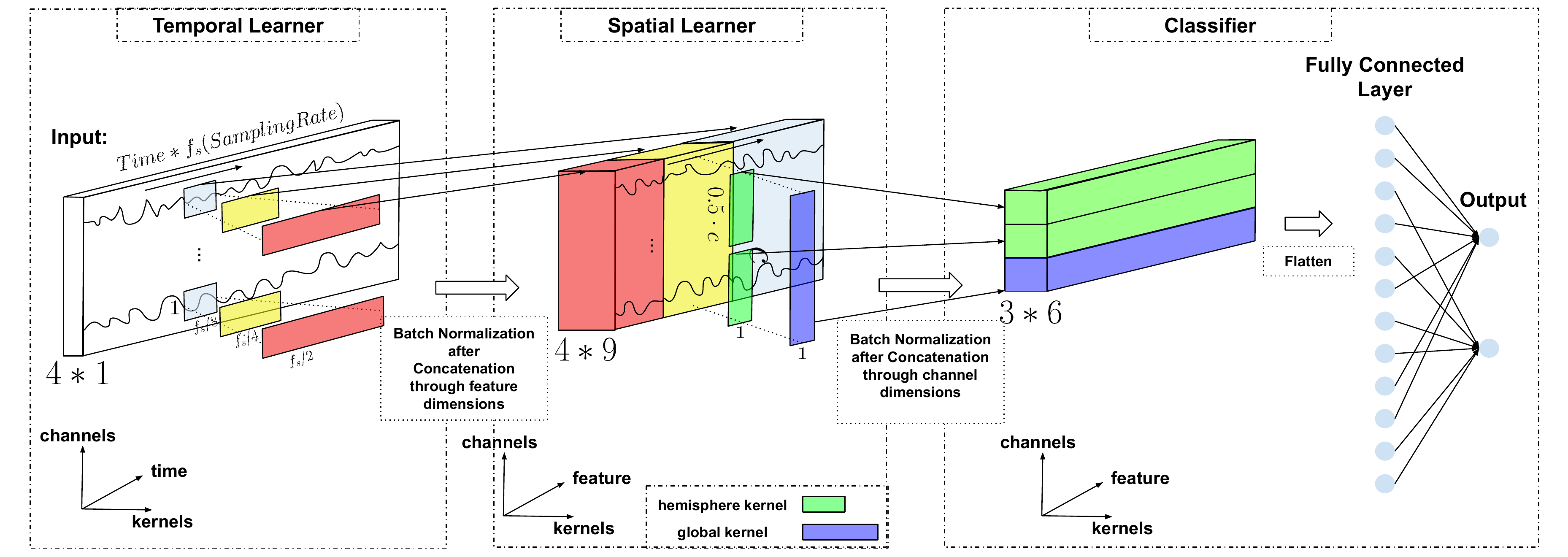}
    \caption{The structure of TSception. The convolution results correspond to the kernels with the same color. TSception can be divided into 3 main parts: temporal learner, spatial learner and classifier. The input is fed into the temporal learner first followed by spatial learner. Finally, the feature vector will be passed through 2 fully connected layer to map it to the corresponding label. The dimension of the input EEG segment is (4 x 1 x 1024) since it has 4 channels, and 1024 data points per channel. There are 9 kernels for each type of temporal kernels in temporal learner, and 6 kernels for each type of spatial kernels in spatial learner. The multi-scale temporal convolutional kernels will operate convolution on the input data parallelly. For each convolution operation, $ReLU(\cdot)$ and average pooling are applied to the feature. The output of each level temporal kernel are concatenated along feature dimension, after which batch normalization is applied. In spatial learner, the global kernel and hemisphere kernel are used to extract spatial information. The output of the two spatial kernels will be concatenated along the channel dimension after $ReLU(\cdot)$ and average pooling. The flattened feature map will be fed into a fully connected layer. After the dropout layer and softmax activation function the classification result will be generated.}
    \label{fig:TS}
\end{figure*}

Temporal Spatial Inception (TSception) can be divided into 3 main parts: temporal learner, spatial learner and classifier. Inspired by Inception block, TSception uses multi-scale convolution kernels in temporal/spatial learners to learn time/space diverse representations simultaneously. Fig.~\ref{fig:TS} shows the general structure of TSception. The input of TSception is the raw EEG signal, making it an end-to-end classification structure. The features are learnt by the temporal and spatial learners automatically. The input is fed into the temporal learner first followed by spatial learner. Finally, the learned feature vector will be passed through 2 fully connected layer to map it to the corresponding label. 
\subsection{Temporal Learner}
The temporal learner consists of multi-scale 1D temporal kernels (T kernels) whose lengths are in different ratios of EEG signal sampling rate $f_{S}$. The ratio coefficients are defined as $ \alpha^{i} \in \mathbb{R}$, where the $i$ is the level of the temporal learner. If the temporal learner has $L$ levels, then i varies from 1 to $L$, and the temporal learner will have $L$ types of temporal kernels. Hence $S_{T}^{i}$, the size of T kernels in $i$ th level, can be defined as:
\begin{equation}
    S_{T}^{i} = \left ( 1, \alpha^{i} \cdot f_{S}\right) 
\end{equation}

From the frequency perspective, in EEGNet, the length of the T kernel is set at half the sampling rate allows for capturing frequency information at 2 Hz and above\cite{Lawhern_2018}.  Emotional states are more related to Alpha (8-15 Hz), Beta (15-32 Hz) and Gamma (>32 Hz)\cite{7946165}, in this work, we expand the temporal perception ranges, letting $L=3, i=1$ to 3 and $\alpha = 0.5$, the ratio coefficients are [0.5, 0.25, 0.125], which can further capture frequency at 4 Hz to above and 8 Hz to above. We hold the hypothesis that by using the multi-scale temporal kernels, the temporal learner can learn multi-frequency representations related to emotional state. From the time perspective, multi-scale T kernels can capture long short-term temporal pattern, providing more diverse representations. The lower level T kernel has a larger ratio coefficient, which gives longer convolutional kernel length and vice versa. The long kernel can learn long term temporal and low-frequency diverse representations. The short kernel extracts short term temporal and high-frequency representations instead. Let $X$ be the raw EEG input segments array. $X = \left [ x^{0} ,x^{1}, ... , x^{n} \right ], x^{n} \in \mathbb{R}^{C \times L}$, where $n$ is the number of EEG segments, $C$ is the number of channels, $L$ is the length of the segments. 
 The multi-scale temporal kernels will operate convolution on the input data parallelly. EEG signal has a low signal-noise ratio, using average pooling can reduce the effect of the noise as well as the feature dimension. After activated by $ReLU(\cdot)$, the feature map is further down-sampled by average pooling. Let $z_{conv}^{i}$ be the output of the $i$ th level temporal kernel, $z_{conv}^{i} \in \mathbb{R}^{B \times T \times C \times F^{i} }$, where the $B$ is the number of samples in each mini-batch, $T$ is the number of each level's T kernel, $C$ is the number of channels, $F^{i}$ is the length of the feature after $i$ th level convolution operation. $z_{conv}^{i}$ is defined as:
 \begin{equation}
     z_{conv}^{i} = AvgPool(ReLU(Conv1D(X, S_{T}^{i})))
 \end{equation}
 where the $S_{T}^{i}$ is the T kernel size, $X$ is the input raw EEG segments array, $Conv1D(\cdot)$ is the 1D convolution operation with the kernel size being $S_{T}^{i}$, step being (1,1). 
 
 The output of each level's T kernel will be concatenated. In order to reduce the internal covariate shift problems in neural networks, batch normalization\cite{ioffe2015batch} is added. Hence the final output of the temporal learner $Z_{T}$, $Z_{T} \in \mathbb{R}^{B \times T \times C \times \sum F^{i}}$ is defined as:
 \begin{equation}
     Z_{T} = f_{bn}([z_{conv}^{0},...,z_{conv}^{i}])
 \end{equation}
 where the $f_{bn}$ is the batch normalization operation, $[\cdot ]$ stands for concatenation operation along the feature (F) dimension. 
\subsection{Spatial Learner}
The spatial learner has multi-scale 1D convolutional kernels whose sizes are related to the location of the EEG channels. There are three types of spatial kernels: global kernel, hemisphere kernel and local kernel. In order to apply three
types kernels, the sequence of channels in the input EEG segments should be carefully arranged. The order of the channels should be $[ channel_{left}, channel_{right}]$, where the $channel_{left}$ are the channels located at left hemisphere, the $channel_{right}$ are the ones on the right hemisphere. Let the input of spatial learner be $X = [x_{0},x_{1}, ... ,x_{n}], x_{n} \in \mathbb{R}^{C \times F}$, where $n$ is the number of EEG segments, $C$ is the number of channels, $F$ is the length of the feature for each channel. 

For the global kernel, it is the same as the ones in EEGNet\cite{Lawhern_2018}, whose size is $(C,1)$, where C is the number of channels. Since the length of the kernel is the same as the channel dimension of the input EEG segment, it can get the global spatial relation pattern. 

Inspired by EEGNet, we further combine the frontal area of brain emotional asymmetry\cite{CRAIG2005566} into the kernel design. The hemisphere kernel is used to extract the relation pattern between left and right hemispheres by sharing the convolutional kernels. The size of the hemisphere kernel is $(0.5 \cdot C, 1)$, and the step is $(0.5 \cdot C, 1)$, where $C$ is the total number of channels. The hemisphere kernel is shared by two hemispheres without overlapping so that the asymmetry pattern can be extracted. 

Further design of the local kernel has the same principle as the above kernels. We can define the sub-area of the brain surface according to functions, and the length of the local kernel would be the number of channels located in the sub-areas. In this work, only the global and hemisphere kernel are used. The local kernel will be considered in future research as a possible alternative.

\subsection{Optimization of TSception}
To optimize the network parameters, we adopt
the back propagation method to iteratively update the
network parameters until the desired criterion is achieved. Cross-entropy cost is used as the objective function. $L1$ Regularization term is added to keep the weights small, making the model simpler and avoiding over-fitting\cite{NIPS2018_7644}. The final loss function is expressed as:
\begin{equation}
    \quad
    L_{\varepsilon}(y, \hat{y}) = L_{Cross-entropy}( y,\hat{y}) + \lambda \sum_{i = 1}^{n}\left | \theta_{i}  \right |
\end{equation}
where the $y$ is ground truth label and the $\hat{y}$ is the predicted label. $\lambda$ is the $L1$ regulation coefficient, $\theta_{i}$ is the $i$ th weight of the model.

To overcome the over-fitting problem, we adopt early stopping in the training process. The stopping criterion is set as the validation accuracy stops increasing for certain epochs. 
\begin{algorithm}
\SetAlgoLined
\KwIn{ Raw EEG data $X$, ground truth label $Y$, model $TSception(\cdot)$, number of temporal kernels $N_T$, number of spatial kernels $N_S$, early stopping patient $p$}

Initialization\;
 $p_{stop} = 0$\;
 $acc_{max} = 0$\;
 \While{$p_{stop}$ < $p$}{
  
  \eIf{$acc_{validation}$ > $acc_{max}$}{
   $acc_{max}$ = $acc_{validation}$\;
   $p_{stop} = 0$\;
   }{
   $p_{stop} += 1$\;
  }
  $\hat{Y} = TSception(X, N_T,N_S)$\;
  $loss = L_{\varepsilon}(Y, \hat{Y})$\;
  $back\_ propagation(loss, optimizer = Adam)$\;
 }
 Save model\;
 \caption{Training Procedure for TSception}
\end{algorithm}

\section{Experiment}
\subsection{Data Acquisition}
In order to study the emotional arousal in an immersive VR environment and evaluate the proposed algorithm, we collected EEG data from 18 healthy subjects (9 Males/9 Females, between the ages of 23-49) using a VR-BCI system. HTC VIVE pro is used as the VR device. Four channels (TP9, AF7, AF8, TP10) EEG data are collected using MUSE EEG headband\cite{8329668}\cite{7862025}. The sampling rate is 256 Hz. The experiments are conducted in an isolated room with soft illumination to avoid external disturbance. Subjects were seated in a comfortable armchair and instructed to avoid undesired movements. The experiment description and the tasks need to be achieved are described to the subject before the experiment. A demo session is added to let the subject be familiar with the system. After the experiment, a survey form will be given to the subject to get the feedback and their emotional state during the experiment.

The system is developed using Unity 3D development platform. There are 2 types of stimuli: low arousal and high arousal.

\begin{figure}[htp]
    \centering
    \includegraphics[width= 8.8cm]{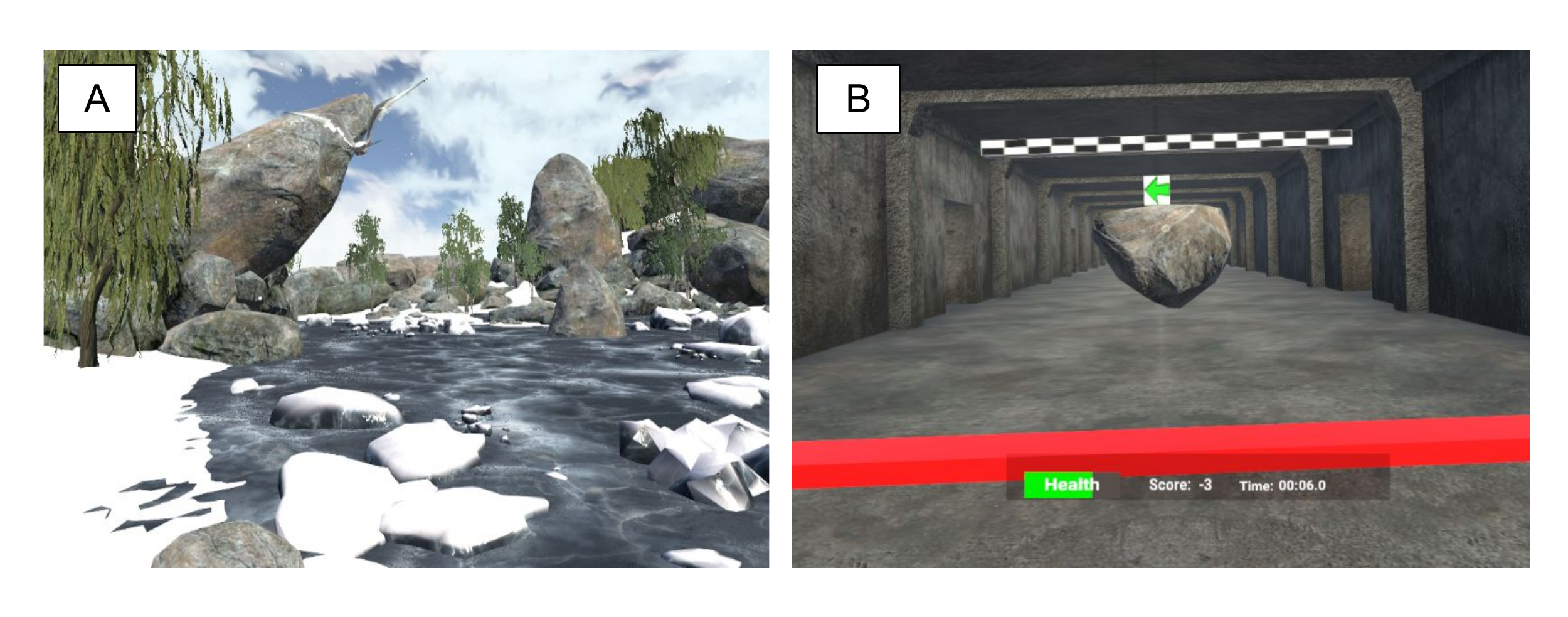}
    \caption{Low arousal stimuli (A) and high arousal stimuli (B). In low arousal stimuli, there is a white bird flies low above the frozen lake in snowing weather, presented in a first-person perspective. For high arousal stimuli, a stone avoiding game is designed to induce high arousal to subject.}
    \label{fig:Scene}
\end{figure}
1) Low arousal stimuli: For this stimulus, as shown in Fig.~\ref{fig:Scene} (A), there is a white bird flies low above the frozen lake in snowing weather, presented in a first-person perspective. The bird flies slow and elegantly, with the soft music played in the background. The design of this stimulus follows Bilgin \textit{et. al.}\cite{8914326}, which indicates that the nature-based and low illumination environment can induce low emotional arousal effectively.

2) High arousal stimuli: Studies \cite{10.1145/2367616.2367629}\cite{10.1016/j.intcom.2009.12.004} show that using complex input (two hands), making subject stressful and increasing difficulty properly can induce higher level arousal to the player. For this stimulus, a stone avoiding game is designed. As seen in Fig.~\ref{fig:Scene} (B), there is a stone coming to the first-person perspective fast and randomly, with the audio effect of moving stone. There will be an arrow appearing when the stone reaches a certain distance from the subject. To avoid the stone, the subject needs to press the left handheld controller button if the arrow points left, and the right handheld controller button if the arrow points right. If the subject presses the correct button, the stone will disappear, the player's score increases by 1. If the wrong button is pressed, the subject will see the stone hits the screen, and the score will be decreased by 1. In order to maintain the arousal level of subjects, adaptive difficulties mechanism\cite{10.1016/j.intcom.2009.12.004} is adopted in the game. The more scores the subject gets, the faster the stone moves, and vice versa.  

\begin{figure}[htp]
    \centering
    \includegraphics[width= 8.8cm]{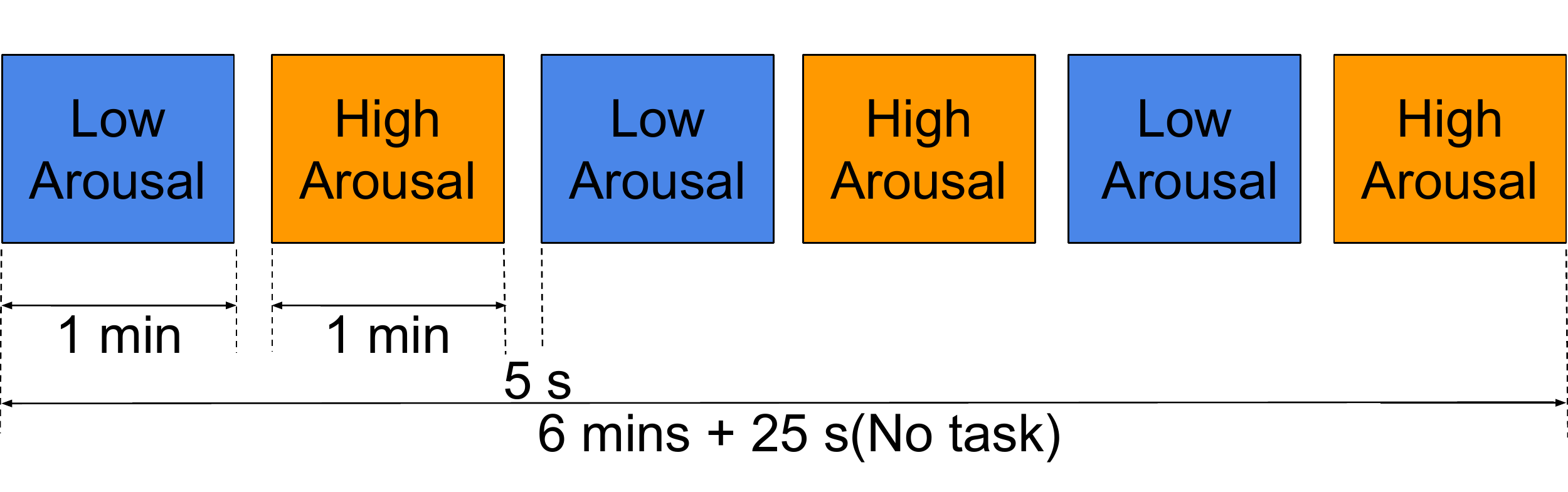}
    \caption{Protocol of the experiment. There are 2 types of stimuli: low arousal and high arousal. For one experiment session, each stimulus lasts for 1 minute, between which there is 5 seconds' relaxing break. Every subject participated in 3 sessions in total.}
    \label{fig:protocal}
\end{figure}

The experiment protocol is shown in Fig.~\ref{fig:protocal}. Every subject participated in 3 sessions in total. There are 2 trials in each session. For one experiment session, each stimulus lasts for 1 minute, between which there is 5 seconds' relaxing break. The TABLE~\ref{Tab:Data} summarizes the experiment information. 

\begin{table}[htbp]
\caption{Experiment Information of Data Acquisition}
\begin{center}
\begin{tabularx}{0.44\textwidth} { 
   >{\centering\arraybackslash}X 
   >{\centering\arraybackslash}X  }
 \hline
\textbf{Factor} & \textbf{Value} \\
\hline
Stimuli& VR scenes and game-playing\\
Number of subjects& 18\\

Number of males& 9 \\

Number of females& 9 \\

Range of age& 23-49 \\

Rating type& Emotional Arousal\\

Rating value& High/Low\\

Channels& TP9, AF7, AF8, TP10\\
Sampling rate& 256 Hz\\
Duration of each subject& 6 mins (3 mins high arousal, 3 mins low arousal)\\
\hline
\end{tabularx}
\label{Tab:Data}
\end{center}
\end{table}

This study, including the data acquisition, was approved by Institutional Review Board of Nanyang Technological University (NTU), Singapore. 
\subsection{Signal Processing}
For the collected data, a band-pass filter from 0.3 Hz to 45 Hz is applied to remove low and high-frequency noise. The electrooculography (EOG) is removed by using MNE open-source python software\cite{GRAMFORT2014446}. The processed data will be used for deep learning methods directly. For SVM, feature extraction is needed. 

The EEG signal is band-pass filtered into multiple frequency bands, using zero-phase Chebyshev Type II filters\cite{4634130}. A total of 9 band-pass filters are used, namely, 4-8 Hz, 8-12 Hz,…, 36-40 Hz. Then the relative power (RP) and DE in 9 frequency bands are used as the features of the SVM input. The RP in $i$ th frequency band can be calculated by:
\begin{equation}
    RP_i = \frac{ \sum x_i^{2}}{\sum_{j}^{n}\sum x_i^{2}}
\end{equation}
where $x_i$ is the data in $i$ th frequency band, $n$ is the number of the frequency bands.

The DE is calculated as\cite{Li2018}:

\begin{equation}
    DE = \frac{1}{2}log2\pi e \sigma^{2}
\end{equation}
where the $e$ is Euler’s constant and $\sigma$ is the standard deviation of $x_i$.

\subsection{Parameter Setting}
The PyTorch library \cite{NIPS2019_9015} is used to implement the proposed model, the source code can be found via $website$ \footnote{https://github.com/deepBrains/TSception} 

Parameters of TSception are selected empirically. There are 3 levels' temporal kernels whose corresponding ratio coefficients are 
$\alpha = \left[0.5, 0.25, 0.125\right]$. For each level, there are 9 convolutional kernels. For the spatial learner, the global and hemisphere kernels are used with 6 convolutional kernels in each type. The hidden node is set as 128 in the first fully connected layers. 

For the training process, Adam optimizer is adopted, with the learning rate being 0.001. The size of the mini-batch is 128. The dropout rate and early stopping patient are 0.3 and 4 respectively. The L1 regulation coefficient $\lambda$ is 1e-06.
\subsection{Experiment Setting}
The subject-dependent experiments are conducted. Since there are 3 sessions for each subject, a "Leave One Session Out" cross-validation strategy is applied. The average accuracy of all subjects and standard deviation are reported as the evaluation criterion. In order to get enough data segments to train the deep learning model better, the raw EEG data is split into 4 seconds' segments by a sliding window, whose moving step is 100 ms (25 data points). Hence 574 samples will be generated per stimuli. For each subject, there are 3 sessions, each session contains 2 stimuli. For leave one session out cross-validation, one session is used as the testing set, another 2 sessions are used as the training set. Among the training set, 80\% of the training set is used for the training process, and the rest 20\% is used as the validation set for early stopping. In one cross-validation step of each subject, the dimension of training set is (1836 x 1 x 4 x 1024), the one for validation set is (460 x 1 x 4 x 1024) and the dimension of testing set is (1148 x 1 x 4 x 1024). The raw data segments will be fed into deep learning methods directly. However for SVM, The manually extracted feature array is used.

\section{Result and Analysis}
The proposed model is compared with EEGNet, LSTM (Using 3 1D CNN layers as feature extractor followed by a 4 layers LSTM) \cite{ZHAO2019312}, SVM using RP as features, SVM using DE as features respectively. For better evaluation of the proposed model, two simplified versions, namely Tception and Sception is added in the comparison. As the names show, Tception is achieved by removing spatial learner of TSception and Sception is the one without temporal learner. 

\begin{table}[htbp]
\caption{Comparison against classification/standard deviation(\%) on subject-dependent experiment with SVM(RP), SVM(DE), EEGNet, LSTM, Tception, Sception, TSception}
\begin{center}
\begin{tabularx}{0.44\textwidth} { 
   >{\arraybackslash}X 
   >{\arraybackslash}X 
   >{\arraybackslash}X  }
 \hline
\textbf{Method} & \textbf{\textit{ACC}}&\textbf{\textit{STD}} \\
\hline
SVM(RP)&80.73  **& 8.51\\

SVM(DE)&82.23  *& 9.07 \\

EEGNet&79.96  *& 11.47\\

LSTM&80.81  **& 8.69\\
\hline
\textbf{Tception}&83.9& 7.42\\

\textbf{Sception}&77.39& 11.47\\

\textbf{TSception}& \textbf{86.03}& 8.99\\

\hline
\end{tabularx}
\begin{tablenotes}
      \small
      \item $p$-value between the method and TSception: * indicating $(p < 0.05)$, ** indicating $(p<0.01)$.
    \end{tablenotes}
\label{Tab:ACC}
\end{center}
\end{table}

\begin{figure}[htp]
    \centering
    \includegraphics[width= 8.8cm]{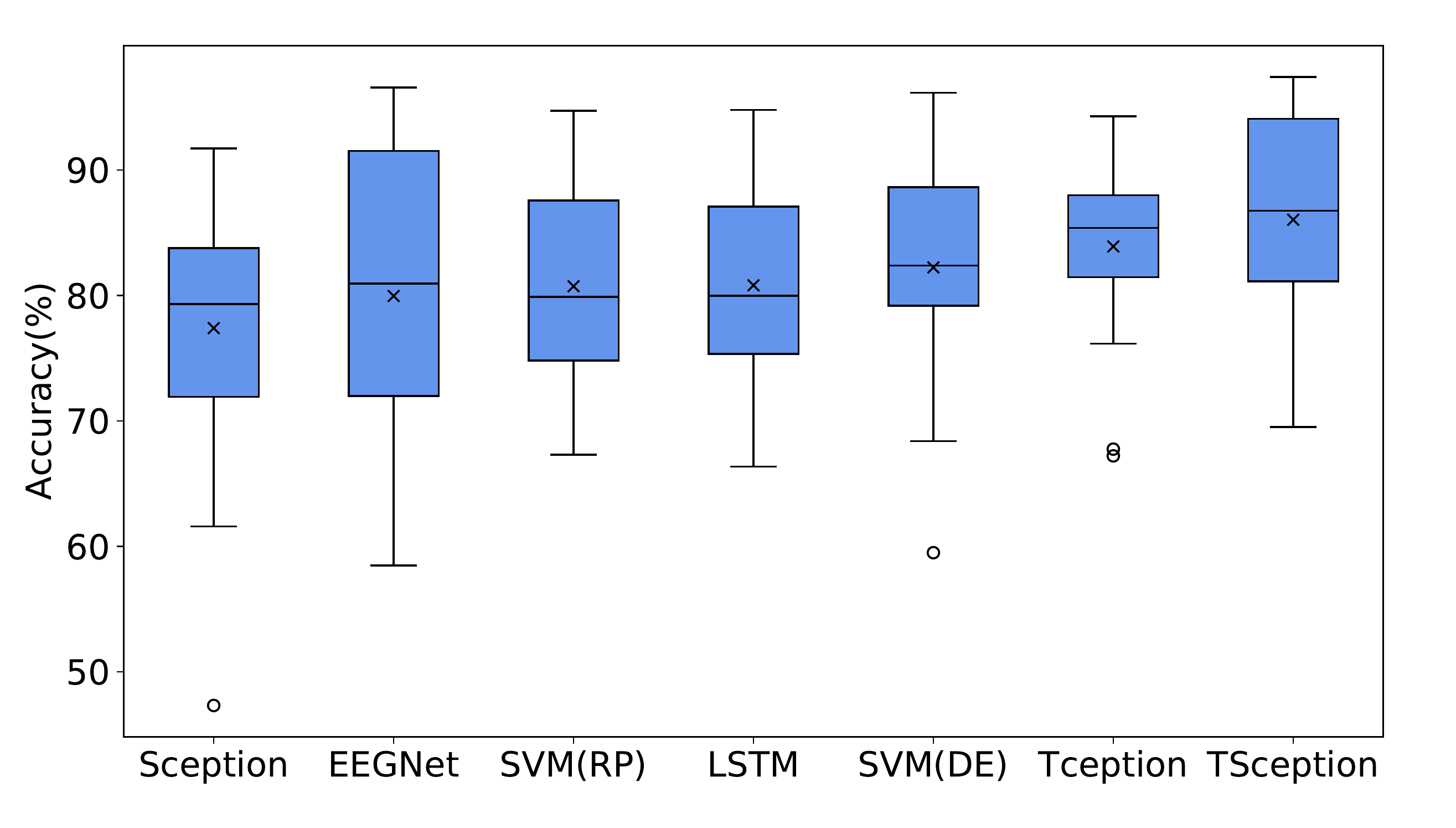}
    \caption{Results of different methods for all subjects. The X-axis is the method, the Y-axis is the classification accuracy(\%). 'x' is the mean of the accuracy for all the subjects. TSception gives the highest accuracy being 86.03\%, followed by Tception (83.9\%) and SVM using DE (82.23\%). The acc of LSTM and SVM using RP are close to each other, being 80.81\% and 80.73\%. The EEGNet is better than Sception with the accuracy being 79.96\% and 77.39\% respectively. }
    \label{fig:Acc}
\end{figure}
TABLE~\ref{Tab:ACC} and Fig. 4 show the classification results for leave one session out cross-validation with SVM(RP), SVM(DE), EEGNet, LSTM, Tception, Sception, TSception.

As shown in the table, DE feature gives higher classification accuracy (82.23\%) than RP feature (80.73\%) for SVM classifier.  For deep learning models, TSception gives the highest accuracy being 86.03\%, followed by Tception (83.9\%) and LSTM (80.81\%). The EEGNet is better than Sception with the accuracy being 79.96\% and 77.39\% respectively. But both EEGNet and Sception give lower accuracy than other deep learning models, indicating more useful patterns are in temporal information than spatial one. EEGNet only has single size temporal kernels and the Sception only extracts the spatial pattern by 1D spatial kernels. Both of them can't extract the dynamic temporal information effectively, even have lower accuracy than SVM with DE feature in 9 frequency bands.

To better understand which part of the TSception contributes more to the classification results, a self-comparison among the TSception and its modified versions is conducted. The detailed structure parameters for TSception and its two simplified versions are shown in TABLE~\ref{Tab:S_acc}. 

\begin{table}[htbp]
\caption{Self-comparison for TSception and two simplified version, Tception and Sception}
\begin{center}
\begin{tabularx}{0.45\textwidth} { 
   >{\centering\arraybackslash}X 
   >{\centering\arraybackslash}X 
   >{\centering\arraybackslash}X 
   >{\centering\arraybackslash}X 
   >{\centering\arraybackslash}X  }
 \hline
\textbf{Method} & Trainable Parameters& Number of temporal kernels &Number of spatial kernels& ACC(\%) \\
\hline
\textbf{Tception}& 822,671 & 9 & - / - & 83.9\\

\textbf{Sception}& 147,902 & - / - & 6 & 77.39\\

\textbf{TSception}& \textbf{53,483} & 9 & 6 & \textbf{86.03}\\
\hline
\end{tabularx}
\label{Tab:S_acc}
\end{center}
\end{table}

As TABLE~\ref{Tab:S_acc} shows, the TSception has less trainable parameters than another 2 models. Compared with Tception and Sception, the final feature vector is much shorter in TSception since it extracts both temporal and spatial patterns in a sequence operation. Hence it dramatically reduces the number of trainable parameters in fully connected layers. Since it extracts the pattern in temporal and spatial information, the classification accuracy is even higher than the other two, which have much more trainable parameters. As for the accuracy of proposed methods, TSception has the highest ACC among the three proposed methods, with 8.34\% improvement over Sception and 2.4\% improvement over Tception. From the results, the temporal learner contributes more than the spatial learner.  Although the Sception gives the lowest accuracy among all the compared methods, the combination of temporal and spatial learner still gives the highest accuracy. There are two possible reasons: 1) the cross information among temporal and spatial helps to improve accuracy; 2) the sequential structure of using two types learners can decrease the parameters of the model significantly, which can overcome over-fitting problems better. 

\section{Conclusion}

In this paper, we proposed TSception, a deep learning framework for EEG emotion classification. It uses multi-scale temporal and spatial convolutional kernels in temporal and spatial learners to learn more discriminative representations in the time and space domain simultaneously. The temporal learner extracts multi-frequency and multi-temporal pattern. The spatial learner takes advantage of the frontal area of brain emotional asymmetry, using hemisphere kernels to extract the information from the right and left hemispheres. 

We collect EEG data from 18 healthy subjects in a VR-BCI system to study the emotional arousal in the immersive VR environment and evaluate the proposed algorithm. Compared with the state of art methods in BCI, such as SVM (RP), SVM (DE), EEGNet, LSTM together with two simple variants of TSception, TSception achieves the highest classification accuracy, being 86.03\%. The proposed model can be applied in EEG signal classification generally due to its general structure. The code of TSception is also made open-access. Exploration for the potential ability of TSception will be included in the future work.

\bibliographystyle{./bibliography/IEEEtran}
\bibliography{./bibliography/mybibliography}

\vspace{12pt}

\end{document}